\documentclass[aps,pre,amsfonts,floatfix, twocolumn,showkeys]{revtex4-1}
\usepackage{amssymb}
\usepackage{amsmath}
\usepackage{amsfonts}
\usepackage{hyperref}
\usepackage{color}
\usepackage{graphicx}
\usepackage{dcolumn}
\usepackage{color}
\usepackage{enumerate}

\begin{document}

\title{
Strategy abundance in  evolutionary many-player games with multiple strategies
}

\author{Chaitanya S. Gokhale}
 \email{gokhale@evolbio.mpg.de}
\author{Arne Traulsen}%
 \email{traulsen@evolbio.mpg.de}
\affiliation{%
Research Group for Evolutionary Theory,\\
Max-Planck-Institute for Evolutionary Biology,\\
August-Thienemann-Stra{\ss}e 2, 24306 Pl\"{o}n, Germany}%

\begin{abstract}
Evolutionary game theory is an abstract and simple, but very powerful way to model evolutionary dynamics.
Even complex biological phenomena can sometimes be abstracted to simple two-player games.
But often, the interaction between several parties determines evolutionary success.
Rather than pair-wise interactions, in this case we must take into account the interactions between many players, which are inherently more complicated than the usual two-player games, but can still yield simple results.
In this manuscript we derive the composition of a many-player multiple strategy system in the mutation-selection equilibrium. 
This results in a simple expression which can be obtained by recursions using  coalescence theory.
This approach can be modified to suit a variety of contexts,
e.g.\ to find the equilibrium frequencies of a finite number of alleles in a polymorphism or that of of different strategies in a social dilemma in a cultural context.
\end{abstract}

\keywords{
many-player game theory, multiple strategies, coalescence theory, abundance}

\maketitle


\section{Introduction}
\label{intro}
Game theory was originally developed in the field of economics to study strategic interactions amongst humans \citep{neumann:1944ef,flood:1952aa}. 
The ``agents" who play against each other have a set of ``strategies" to choose from.
The payoff which an agent gets depends on its own strategy and the strategy of the opponent.
A player can decide which strategy to play against an opponent of a given strategy.

In evolutionary game theory players are born with fixed strategies instead, \citep{maynard-smith:1982to}
which are  considered to be inherited traits.
As usual, we assume a population game in which every player effectively plays against the average opponent.
The success of a strategy depends on the number of players of that strategy and also the number of players with other strategies.
A classical example is the Lotka-Volterra equation \citep{lotka:1910aa,volterra:1928aa,hofbauer:1998mm}.
If the number of wolves increases then the number of hares will decrease in turn leading to a decrease in the number of wolves.
Evolutionary game dynamics studies the change in the frequencies of the strategies \citep{nowak:2006bo}, which depends on mutation, selection and drift.

A recurrent and obvious question asked in the study of games is which is the best strategy?
Assuming an infinitely large population we can approach this question by the traditional replicator dynamics \citep{hofbauer:1998mm}.
The frequency of a strategy will increase if its average payoff is greater than the average payoff of the whole population.
That is, if the individuals of a particular strategy are doing better on average than the individuals of other strategies then that strategy spreads.
The average payoff of a strategy is also dependent on the frequency of the strategy.
For finite populations one must resort to stochastic descriptions \citep{ficici:2000aa,schreiber:2001aa,nowak:2004pw}.
One important quantity is the fixation probability.
Consider two strategies $A$ and $B$ in a population of size $N$.
Let the population be almost homogenous for $B$ with only a single $A$.
If there is no fitness difference amongst the strategies, i.e. selection is neutral, then the probability that the $A$ individual will take over the entire population is $1/N$.
If this probability is greater than $1/N$ we say that strategy $A$ is favoured by selection.
When there are multiple strategies in the population, then a pair-wise comparison between the fixation probabilities of all the strategies will reveal which is the most abundant strategy 
\citep{fudenberg:2006ee,hauert:2007aa,hauert:2008bb,van-segbroeck:2009mi,sigmund:2010aa}.
This analysis requires the assumption of low mutation rates.

When mutations become more frequent then the concept of fixation itself is problematic and hence also that of fixation probability.
In such a case we resort to the average frequency of a strategy which is maintained at the mutation-selection balance.
This has been termed as the abundance of a strategy \citep{antal:2009hc}.

Consider $n$ strategies which are effectively neutral against each other.
In such a case the abundance of all the strategies in the stationary state will be just $1/n$.
Usually there are fitness differences between the strategies.
If the abundance of a strategy is greater than that of all the other strategies then we can say it is favoured under the effects of mutation, selection and drift.
Hence for $n$ strategies, the $k^{th}$ strategy will be favoured if the abundance of $k$ is greater than $1/n$.
Calculating the abundance of a strategy is a non-trivial exercise even when assuming weak selection.
\cite{antal:2009hc} have developed such an approach based on coalescence theory for the case of two-player games and $n$ strategies.
Under certain conditions and weak selection, one can calculate the most abundant strategy for arbitrary mutation rates even in structured populations \citep{antal:2009aa,tarnita:2009df,tarnita:2009jx} and bimatrix games \citep{ohtsuki:2010aa}.

Usually two-player interactions are studied in evolutionary game theory.
The analysis of Antal et al. is also for two-player games.
The interactions which we usually use as examples in evolutionary game theory are in general multi-player interactions making the systems nonlinear \citep{nowak:2010na}.
A classical example where a certain minimum number of individuals are required to complete a task is group hunting.
\cite{stander:1992aa} studied cooperative hunting in lions.
A typical hunting strategy is to approach the prey from at least three sides to cutoff  possible escape paths.
This hunting approach is impossible with only two hunters, i.e.\ a two-player game theoretic approach would be insufficient to capture the dynamics.
Although evolutionary dynamics of multi-player games has received growing interest in the recent years, the main focus has been the Public Goods Game and its variants \citep{hauert:2002te,milinski:2006aa,rockenbach:2006aa,hauert:2007aa,santos:2008xr,pacheco:2009aa,souza:2009aa,veelen:2009ma,archetti:2011aa}.
Only few authors consider general evolutionary many person games  
 \citep{hauert:2006fd,kurokawa:2009aa,gokhale:2010pn}.
We extend the approach developed by \cite{antal:2009hc} for two-player games and multiple strategies to multi-player games.
We show that in the limit of weak selection it is possible to calculate analytical results for $n$ strategies and $d$ players for arbitrary mutation rates.
For a three-player game the mathematical analysis is described in detail.
It is followed by an example with simulations supporting the analytical result.
Lastly we discuss how the methodology can be extended for $d$-player games and argue that a general approach is possible, but tedious.

\section{Abundances in the stationary state for three-player games}
\label{model}

\cite{antal:2009hc} have developed an approach to find the abundances of $n$ strategies in a two-player game ($d=2$).
For a two-player game even with $n$ strategies, the payoff values can be represented in the usual payoff matrix form.
They can be represented as quantities with two indices, $a_{k,h}$.
We increase the complexity first by adding one more player ($d=3$).
This adds another index for the third player's strategy set, $a_{k,h,i}$.
To calculate the average change in the frequency of a strategy we thus need to take into account this payoff `tensor'.

We calculate the abundance of a strategy at the mutation-selection equilibrium.
We begin by checking if there is a change in the frequency of a strategy, say $k$ on average, due to selection.
The average change under weak selection is given by
\begin{eqnarray}
\label{replike}
\langle \Delta x_k^{sel} \rangle_\delta = \frac{\delta}{N} \left(\sum_{h,i} a_{k,h,i}\langle x_k x_h x_i\rangle - \sum_{h,i,j} a_{h,i,j}\langle x_k x_h x_i x_j\rangle \right),\nonumber \\
\end{eqnarray}
where the angular brackets denote the average in the neutral stationary state.
The $\delta$ (selection intensity) as a lower index on the left hand side, however, denotes that the average is obtained under (weak) selection.
If we pick three individuals in the neutral stationary state, then the probability of the first one to have strategy $k$, the next one $h$ and the last $i$, is given by the angular brackets in the first sum, $\langle x_k x_h x_i\rangle$.
Furthermore, $a_{k,h,i}$ denotes the payoff values obtained by a strategy $k$ player when pitted against two other players of strategy $h$ and $i$.
For $n$ strategies the sums run from $1$ to $n$.
This equation is the special case of a $d=3$ player game.
The derivation for arbitrary $d$ is given in \ref{eq1app}.
The above equation is similar to the replicator equation, which is also based on the difference between the average payoff of a strategy and the average payoff of the population, but as we will see below, here the averages on the right hand side also include mutations.

To incorporate mutations in the process, we write the total expected change due to mutation and selection as
\begin{eqnarray}
\label{xtot}
\Delta x ^{tot}_k = \Delta x^{sel}_k (1-u)+ \frac{u}{N}\left(\frac{1}{n} - x_k\right).
\end{eqnarray}
The first term is the change in the frequency in the absence of mutation.
In presence of mutations, the second term shows that the frequency can increase by $1/(nN)$ by random mutation and decrease by $x_k/N$ due to random death.
A mutation means that with a certain probability $u$, the strategy $k$ can mutate to any of the $n$ strategies.

We are interested in the abundance of a strategy in the stationary state.
In the stationary state, the average change in frequency is zero, $\langle \Delta x ^{tot}_k \rangle_\delta = 0$, as the mutations are balanced by selection.
Averaging Eq.\ \ref{xtot} under weak selection thus gives us
\begin{eqnarray}
\label{abundeq}
\langle x_k \rangle_\delta = \frac{1}{n} + N \frac{1-u}{u} \langle \Delta x^{sel}_k \rangle_\delta.
\end{eqnarray}
This is our quantity of interest, the abundance of a strategy when the system has reached the stationary state.
For $d=2$ player games, this quantity is given by \cite{antal:2009hc}.
For the abundance of a strategy to be greater than neutral, $\langle x_k \rangle_\delta > \frac{1}{n}$, the change in frequency in the stationary state due to selection must be greater than zero, $\langle \Delta x^{sel}_k \rangle_\delta >0$.

Thus, we need to resolve the right hand side of Equation \ \ref{replike}.
Consider the first term in the brackets.
In the neutral stationary state the number of combinations in the sums reduces due to symmetry, e.g. $\langle x_i x_j x_j \rangle = \langle x_j x_i x_j \rangle = \langle x_j x_j x_i \rangle$.
Hence, we need to calculate only three different terms, $ \langle x_1 x_1 x_1 \rangle $, $ \langle x_1 x_2 x_2 \rangle $ and $ \langle x_1 x_2 x_3 \rangle $.
Also for $d$ player games, the terms in the sums are reduced.
For the second term in the brackets we need to calculate five different types of averages, $\langle x_1 x_1 x_1 x_1\rangle$, $\langle x_1 x_2 x_2 x_2\rangle$, $\langle x_1 x_1 x_2 x_2\rangle$, $\langle x_1 x_1 x_2 x_3\rangle$ and $\langle x_1 x_2 x_3 x_4\rangle$.
These averages are derived in the \ref{averagesapp}.
Using an approach from coalescence theory, we derive $s_i$, the probability that $i$ individuals chosen from the stationary state all have the same strategy.
Hence $s_4$ is the probability that four individuals chosen in the stationary state all have the same strategy.
If there are in all $n$ strategies, then the probability that all have exactly strategy $1$ is $s_4/n$.
Hence, $\langle x_1 x_1 x_1 x_1 \rangle = \langle x_2 x_2 x_2 x_2 \rangle = \ldots = \langle x_n x_n x_n x_n \rangle = s_4/n$.
Conversely, $\bar{s}_i$ is the probability that if we choose $i$ individuals in the stationary state, each has a unique strategy.
Knowing these averages helps us resolve Eq.\ \eqref{replike},
\begin{eqnarray}
\langle \Delta x_k^{sel} \rangle_\delta
= \frac{\delta \mu (L_k + M_k \mu + H_k \mu^2)}{N n (1+\mu) (2+\mu) (3+\mu)} 
\end{eqnarray}
where $\mu = N u$ and $L_k$, $M_k$ and $H_k$ are functions consisting only of the number of  strategies $n$ and the payoff values $a_{k,h,i}$ (see \ref{averagesapp}).
Using this and evaluating Eq.\ \eqref{abundeq} gives us the abundance of the $k^{th}$ strategy.
\begin{eqnarray}
\label{finalres}
\langle x_k \rangle_\delta = \frac{1}{n}\left[ 1 + \frac{ \delta (N- \mu) (L_k + M_k \mu + H_k \mu^2) }{ (1+ \mu) (2+ \mu) (3+ \mu)}\right].
\end{eqnarray}
This expression is valid for large population sizes, $N\delta \ll 1$ and any $\mu =N u$.
Since we have $0\leq u \leq 1$, $\mu$ is bounded by $0\leq \mu \leq N$.

We arrive at the result with an implicit assumption that there are at least four strategies.
For $n \leq d$, each player cannot have a unique strategy and hence we need to set the corresponding terms to zero (see \ref{averagesapp}).
If there are less than $n=4$ strategies then $\bar{s}_4$ vanishes.
This does not affect our general result as the affected terms in $L_k$, $M_k$ and $H_k$ simply vanish.

\section{An example for three-player games with three strategies}
\label{example}

To illustrate the analytical approach we explore an evolutionary three-player game with three strategies $A$, $B$ and $C$.
Let our focal individual play strategy $A$.
The other two players can play any of the three strategies.
This can lead to a potential complication.
Consider the combinations $AAB$ or $ABA$.
If the order of players does not matter, then both these configurations give the same payoffs but if they do matter then we need to consider them separately.
Here we assume random matching, and hence the order drops out (e.g. $AAB$ and $ABA$ are equally likely).
We consider an arbitrary game as denoted in Table \ref{paytab}.

\begin{figure}
\includegraphics[width=\columnwidth]{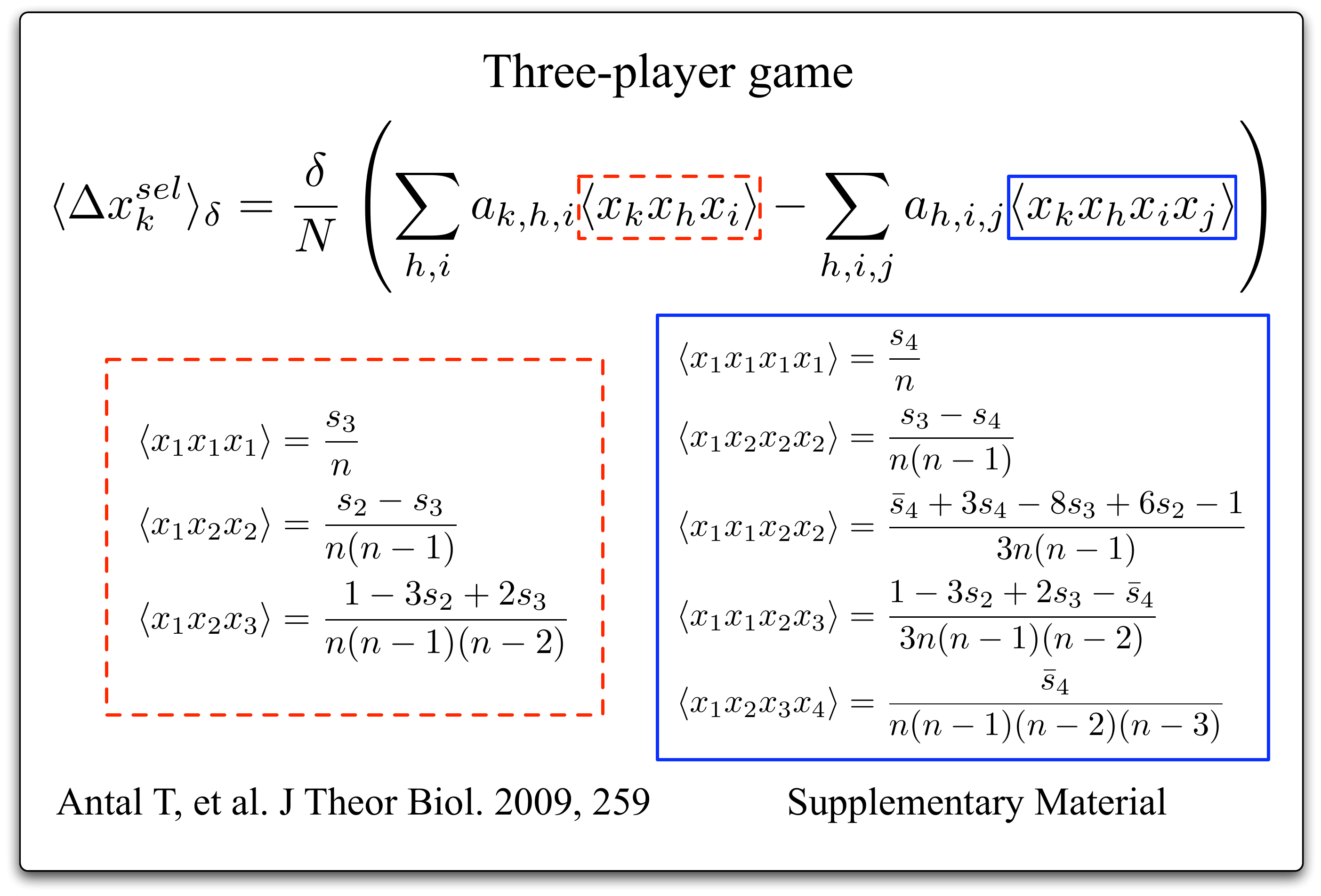}
\caption{The average change in the frequency of strategy $k$ due to selection, $\langle \Delta x_k^{sel} \rangle_\delta$ for a three-player game.
Notice first the similarity to the replicator equation where also we look at how a strategy is faring compared to the population.
The first term in the bracket is analogous to the average fitness of strategy $k$.
If we pick three individuals in the stationary state, then the probability that the first one has strategy $k$, second $h$ and the third $i$ is given by $\langle x_k x_h x_i\rangle$  (dashed box).
Even for $n$ strategies there are only three possible combinations, either all can have the same strategy, a pair has the same strategy or all three have different strategies.
These probabilities were calculated by \cite{antal:2009hc}.
The $s_i$'s appearing in the averages are the probabilities that if we choose $i$ individuals from the stationary distribution then they all have the same strategy.
The second term in the bracket is analogous to the average fitness of the population in the stationary state.
For this we need to pick four individuals and look for all the different combinations (solid box).
For $n$ strategies, five combinations can explain all the different configurations.
These range from all the individuals having the same strategy $\langle x_1 x_1 x_1 x_1 \rangle$ to all having a different strategy $\langle x_1 x_2 x_3 x_4 \rangle$ (\ref{averagesapp}).
For the latter, we calculate $\bar{s}_i$, the probability that we choose $i$  individuals from the stationary distribution and each of them has a unique strategy.
For a general $d$-player game we need to pick $d$ individuals for the first term and $d+1$ for the second.}
\label{fig:1}
\end{figure}

We need to calculate the average change in the frequency of strategy $A$ due to selection, i.e. Eq.\ \eqref{replike}.
We denote the co-efficients of the averages in the first sum by $\alpha_1$, $\alpha_2$ and $\alpha_3$.
Hence for example, $\alpha_3 = a_{A,B,C} + a_{A,C,B}$.
Similarly for the second sum we have $\beta_1$ to $\beta_4$ (Note that $\beta_1 = \alpha_1 = a_{A,A,A}$).
Thus we have,
\begin{eqnarray}
\sum_{h,i} a_{A,h,i}  \langle x_A x_h x_i\rangle &=& \alpha_1 \langle x_A x_A x_A \rangle + \alpha_2  \langle x_A x_B x_B \rangle \nonumber \\
&&+ \alpha_3 \langle x_A x_B x_C \rangle
\end{eqnarray}
and
\begin{eqnarray}
\sum_{h,i,j} a_{h,i,j}\langle x_A x_h x_i x_j\rangle  &=& \beta_1 \langle x_A x_A x_A x_A\rangle + \beta_2 \langle x_A x_B x_B x_B\rangle \nonumber \\
&&+ \beta_3 \langle x_A x_A x_B x_B\rangle + \beta_4 \langle x_A x_A x_B x_C\rangle.\nonumber \\
\end{eqnarray}
Note that the term $\langle x_A x_B x_C x_D\rangle$ which would appear with a factor $\beta_5$, does not appear, as we have only three strategies and thus $\bar{s}_4 = 0$.
This also alters the definition of  $\langle x_A x_A x_B x_C\rangle$ and $\langle x_A x_A x_B x_B\rangle$ (see Figure \ref{fig:1}, all terms dependent on $\bar{s}_4$ are affected).

\begin{figure}
\includegraphics[width=\columnwidth]{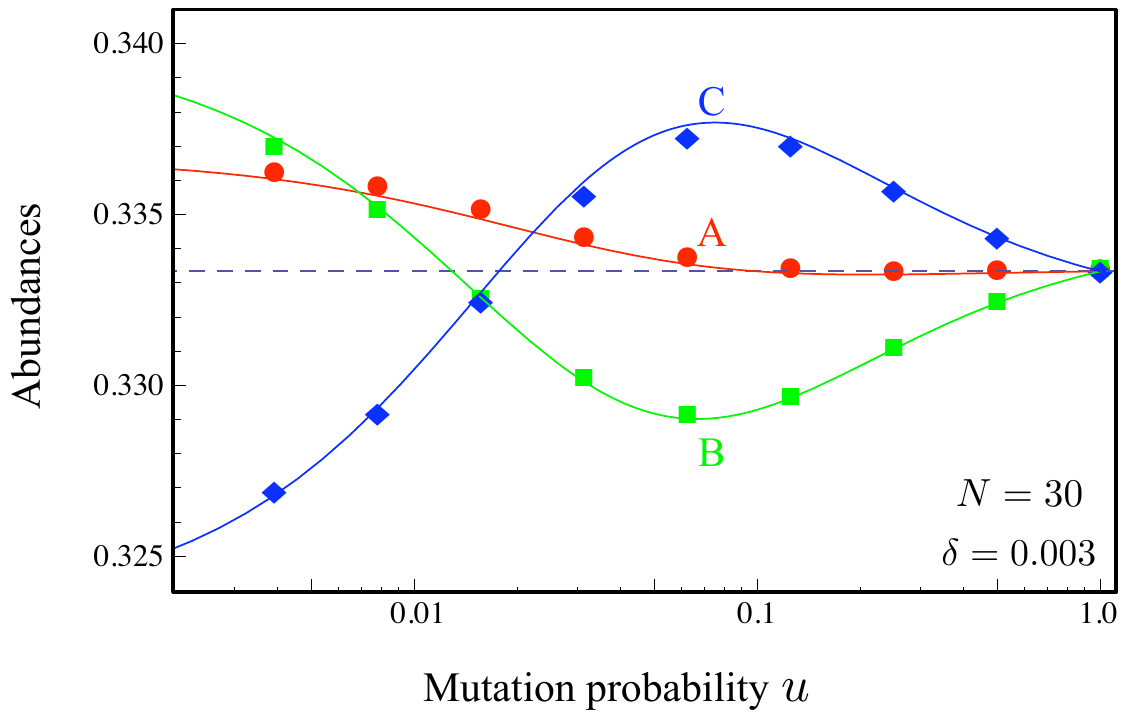}
\caption{
For a three-player game with three strategies ($d=3;n=3$) we plot the average abundances of the three strategies as a function of the mutation probability $u$.
The payoff table from Table \ref{paytab} is used.
The lines are the solutions of Eq.\ \eqref{abundeq} and the symbols are the simulation results for the three strategies.
Although the calculations are valid for large populations we see a good agreement even for a population size of  $N=30$
(selection intensity $\delta=0.003$, simulation points are obtained 
averaging over $20000$ independent runs, each over $2 \times 10^6$ time steps after a transient phase of $N$ time steps).
}
\label{fig:2}
\end{figure}

We know the form of $L_k$, $M_k$ and $H_k$ from \ref{averagesapp} as,
\begin{eqnarray}
L_k  &=& \tfrac{1}{n} \left[2 \alpha_1 (n-1) + 3 \alpha_2 - 2 \beta_2 - \beta_3\right] \\
M_k &=& \tfrac{1}{n^2} \left[\left(3 n-3\right) \alpha_1 +\left(n+3\right) \alpha_2 +3 \alpha_3 - 3 \beta_2 - 2 \beta_3 - \beta_4\right] \nonumber \\
\\
H_k &=& \tfrac{1}{n^3} \left[n (\alpha_1 + \alpha_2 + \alpha_3) - (\beta_1 + \beta_2  + \beta_3  + \beta_4  + \beta_5)\right]
\end{eqnarray}
With $L_k$, $M_k$ and $H_k$ as above, Eq.\ \eqref{finalres} for $n=3$ reduces to,
\begin{eqnarray}
\langle x_A \rangle_\delta = \frac{1}{3}\left[ 1 + \frac{\delta (N- \mu) (L_k + M_k \mu + H_k \mu^2) }{ (1+ \mu) (2+ \mu) (3+ \mu)}\right].
\end{eqnarray}
This gives us the abundance of strategy $A$ at the mutation selection equilibrium.
Repeating the same procedure for strategies $B$ and $C$ gives the analytical lines in 
Figure \ref{fig:2}.
Although the analytical solutions are valid for large population sizes only, we still see a good agreement between the simulation and theory results, even for a population size as small as $30$.

\begin{table}
\begin{center}
\begin{tabular}{ccccccc}
\hline\hline
\hspace{0.2cm}
\begin{minipage}[c]{2cm}
\begin{center}
\vspace{0.1cm}
Weights
 \\ 
( Total 9 )
\vspace{.2cm}
\end{center}
\end{minipage} & 1	& 2 & 2 & 1 & 2 & 1	\\
\hline
 & AA & AB & AC & BB & BC & CC	\\
 \hline
 A 	& 2 & 2  & 3 & 1 & 3 & 4	\\
 B 	& 2 & 1 & 2 & 3	& 0 & 2\\
 C 	& 2 & 12 & 2 & 0 & 1 & 3	\\
\hline\hline
\end{tabular}
\caption{An example payoff table for $d = 3$ and $n=3$.
Consider a three-player game with three strategies $A$, $B$ and $C$.
The strategy of the focal individual is in the column on the left.
For example the payoff received by a $C$ individual when playing in a configuration of $CAB$ is  $12$.
From the focal individual's point of view there are two ways of this configuration $CAB$ and $CBA$ as it is twice as likely as compared to e.g. $CAA$.
Hence we weight that payoff value by $2$ when calculating the average payoff of strategy $C$.}
\label{paytab}
\end{center}
\end{table}

\section{Abundances in $\mathbf{d>3}$ player games.}

We can repeat the whole procedure for $d=4$ player games with $n$ strategies.
The formula for the abundance remains the same, Eq.\ \eqref{abundeq}, but the average change due to selection, Eq.\ \eqref{replike}, becomes more complicated.
We need to add an index in the sums,
\begin{eqnarray}
\langle \Delta x_k^{sel} \rangle_\delta &=& \frac{\delta}{N} \Big(\sum_{l,m,n} a_{k,l,m,n}\langle x_k x_l x_m x_n\rangle \nonumber \\
&&- \sum_{l,m,n,o} a_{l,m,n,o} \langle x_k x_l x_m x_n x_o\rangle \Big)
\end{eqnarray}
The first term is comparatively simple as we already know all the different ways of picking four individuals.
For the second term we need to know the different possible combinations of strategies when picking five individuals from the neutral stationary state.

For $d$ players and $n$ strategies we can construct an expression analogous to Eq.\ \ref{replike}.
Consider for example the strategies played by $d$ individuals denoted by, $r_1, r_2, r_3 \ldots r_d$.
Note that each of these can be a strategy from the strategy set $1,2,3 \ldots n$.
Let $k$ be our strategy of interest.
Then the expression for the change of strategy $k$ due to selection is given by,
\begin{eqnarray}
\label{dplayers}
\langle \Delta x_{k}^{sel} \rangle_\delta &=& \frac{\delta}{N}  \Bigg( \sum_{r_2,\ldots r_d} a_{k,r_2,\ldots r_d } \langle x_k x_{r_2} x_{r_3} \ldots x_{r_d} \rangle  \nonumber \\
\!\!\!\! &&- \sum_{r_2,\ldots r_{d+1}} a_{r_2,\ldots r_{d+1} } \langle x_k x_{r_2} x_{r_3} \ldots x_{r_{d+1}} \rangle \Bigg) \nonumber \\
\end{eqnarray}
where the sums range as usual from $1$ to $n$ (\ref{eq1app}).
Solving this and plugging it in Eq.\ \eqref{abundeq} gives the generalized expression for the abundance of strategy $k$ for an $n$ strategy, $d$-player game.
We see that in the first sum the averages are for choosing $d$ players but for the second it is $d+1$.
Hence we need to calculate the probabilities of the form $s_{d+1}$, but $s_{d+1}$ depends on $s_d$.
Thus we have to solve the expression recursively e.g.
for $d=6$, we will need to pick $7$ players at most and we must solve the expression for $d = 2,3,4,5,6$ before ($d=2$ has been solved by \cite{antal:2009hc} and $d=3$ in this paper).
As $d$ increases calculating $s_{d+1}$ is not enough and we will also need to calculate terms such as $\bar{s}_{d+1}$ which is already the case for $d=3$.

\section{Special case: Two strategies, $\mathbf{n=2}$}

Games with two strategies have been very well studied.
In two-player games with two strategies, one strategy can replace another with a higher probability if the sum of its payoff values is greater than the sum of the payoff values of the other strategy.
This is valid under small mutation rates for deterministic evolutionary dynamics \citep{kandori:1993aa}.
The result also holds for for different dynamical regimes under specific limits of selection intensity and mutation rates \citep{fudenberg:1992bv,nowak:2004pw,antal:2009th}.
Recently it has been shown that this result can be generalized for $d$-player games with two strategies \citep{kurokawa:2009aa,gokhale:2010pn}.

Hence the condition which we find for $d$-player games should be identical to $L_k > 0$ derived in this paper for $d$ players.
We check this for $d=3$,
\begin{eqnarray}
L_k  &=& \tfrac{1}{2} \left[2 \alpha_1 (2-1) + 3 \alpha_2 - 2 \beta_2 - \beta_3 \right]
\\
\nonumber 
&=&\tfrac{1}{2}  [ 2 a_{1,1,1} + 3 ( a_{1,1,2} + a_{1,2,1} + a_{1,2,2} )\nonumber \\ 
&& - 2 ( a_{1,1,2} + a_{1,2,1} + a_{2,1,1} + a_{2,2,2} ) - a_{1,2,2} - a_{2,1,2} - a_{2,2,1} ] \nonumber \\
\end{eqnarray}
Thus $L_k > 0$ is equivalent to,
\begin{eqnarray}
2 a_{1,1,1} +&&   a_{1,1,2} + a_{1,2,1} +  2 a_{1,2,2} \nonumber \\
&&> 2 a_{2,1,1} +   a_{2,1,2} + a_{2,2,1} +  2 a_{2,2,2} \nonumber \\
\end{eqnarray}
If we assume that the order of players does not matter then we have $a_{1,1,2} = a_{1,2,1}$ and $a_{2,1,2} = a_{2,2,1}$.
This yields
\begin{eqnarray}
a_{1,1,1} +   a_{1,1,2} +  a_{1,2,2} >  a_{2,1,1} +   a_{2,1,2} +  a_{2,2,2},
\end{eqnarray}
which is exactly the condition that has been obtained previously using different methods and different notation by \cite{kurokawa:2009aa} and \citep{gokhale:2010pn}.
\section{Application to a task allocation problem}
To demonstrate the power of the approach we are motivated by \cite{wahl:2002aa} who studies the problem of task allocation and of the evolution of division of labour.
\cite{wahl:2002aa} studied a two-player game between task $1$ specialists ($T_1$), task $2$ specialists ($T_2$) and generalists. 
Instead, we have $T_1$, $T_2$ and freeloaders $F$.
We can think of this problem  in context of bacteria that need two types of enzymes to obtain resources from the environment.
One strain produces one type of enzyme at a cost $c_1$ and 
another strain produces the second type of enzyme at a cost of $c_2$.
We also have the freeloading strain which does not produce any enzyme but can get resources by the help of the other two strains.
The benefit of getting 
the resources is given by $b$.
We have the condition that the total cost is less than the benefit accrued i.e. $b> c_1 + c_2$.
Further assume that our contenders are conservative in the enzyme production.
Sensing who they are pitted against, the strains share the costs of producing the enzyme.
Thus a two-player payoff matrix for such a setting can be written down as in Table \ref{2plpaytabex}.
It is hard to imagine though that the bacteria interact only in a pair-wise fashion.
Although it is hard to judge how many players are interacting, we can at least increase the complexity by one more player and study what effect this has on the abundances of the strains.
Therefore, we study the system
in a three-player setting.
In this case the payoff table will look like \ref{3plpaytabex}.
\begin{table}
\begin{center}
\begin{tabular}{ccccccc}
\hline\hline
 & $T_1$ & $T_2$ & $F$ 	\\
 \hline
$T_1$ & $\frac{-c_1}{2}$ & $b-c_1$  & $-c_1$ 	\\
$T_2$& $b-c_2$ & $\frac{-c_2}{2}$ & $-c_2$ \\
$F$ 	& 0 & 0 & 0 \\
\hline\hline
\end{tabular}
\caption{
Payoff table for a two-player game with three strategies, $T_1$ i.e. specialising in task $1$ or $T_2$ i.e. specialising in task $2$.
$T_1$ produces
enzyme $1$ and 
$T_2$ produces 
enzyme $2$.
When both the enzymes are present then a benefit $b$ is obtained.
The cost of producing enzyme $1$ is $c_1$ and the cost to produce enzyme $2$ is $c_2$.
If the partner in the game is of the same strategy then the cost is shared.
The freeloading strategy $F$ does not pay any cost and so the benefits of the resource are unavailable to it.}
\label{2plpaytabex}
\end{center}
\end{table}
\begin{table}
\begin{center}
\begin{tabular}{ccccccc}
\hline\hline
\hspace{0.2cm}
\begin{minipage}[c]{2cm}
\begin{center}
\vspace{0.1cm}
Weights
 \\ 
( Total 9 )
\vspace{.2cm}
\end{center}
\end{minipage} & 1	& 2 & 2 & 1 & 2 & 1	\\
\hline
 & $T_1 T_1$ & $T_1 T_2$ & $T_1 F$ & $T_2 T_2$ & $T_2 F$ & $F F$	\\
 \hline
 $T_1$ 	& $\frac{-c_1}{3}$ & $b-\frac{c_1}{2}$  & $\frac{-c_1}{2}$ & $b-c_1$ & $b-c_1$ & $-c_1$	\\
 $T_2$ 	& $b-c_2$ & $b-\frac{c_2}{2}$ & $b-c_2$ & $\frac{-c_2}{3}$	& $\frac{-c_2}{2}$ & $-c_2$\\
 $F$ 	& 0 & b & 0 & 0 & 0 & 0	\\
\hline\hline
\end{tabular}
\caption{
Payoff table for the same game as discussed in Table \ref{2plpaytabex}.
$T_1$ and $T_2$ refer to specialising in task $1$ and $2$ namely producing enzyme $1$ and $2$.
Note the costs can also be shared if at least one of the game partners is of the same strategy.
In this case the freeloaders can get the benefit when the other two players produce both enzymes.
}
\label{3plpaytabex}
\end{center}
\end{table}

\begin{figure*}
\includegraphics[width=2.0\columnwidth]{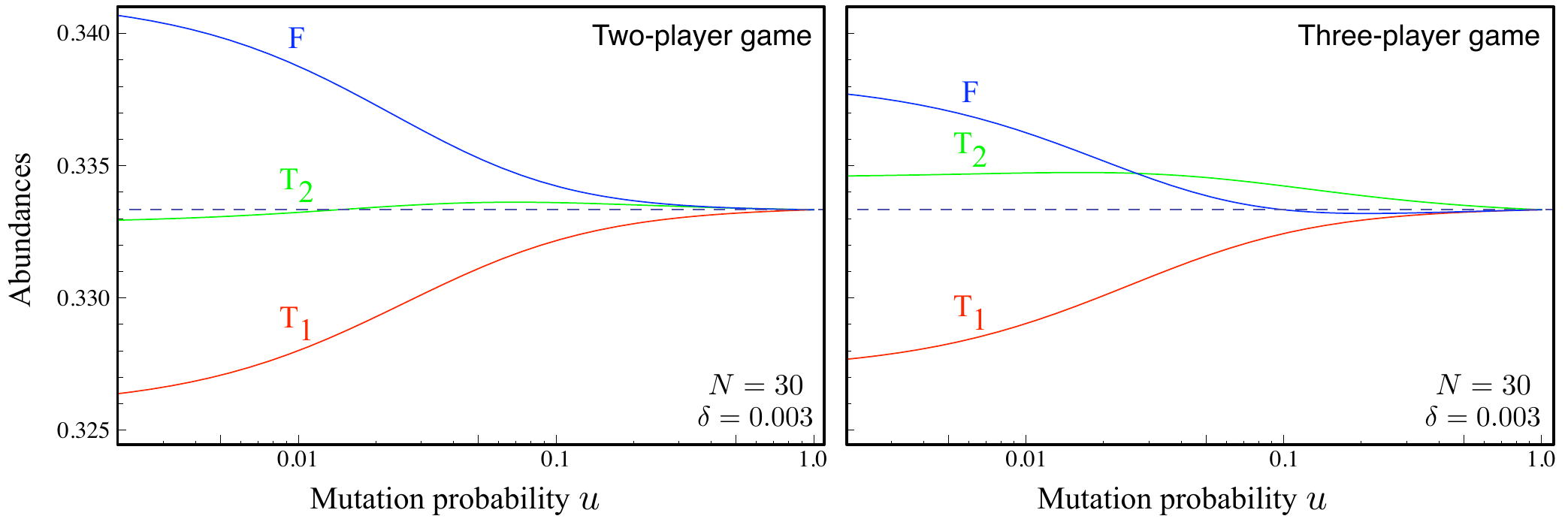}
\caption{
The strains of a bacteria $T_1$ and $T_2$ produce the enzymes $1$ and $2$ which when present together provide a benefit $b$.
When more than one individual of the same strain is present then the production costs for the enzymes, $c_1$ and $c_2$ are shared.
A third strain $F$ does not produce any enzyme and thus avoids the costs and cannot obtain the benefit on its own.
We can analyse the interactions by assuming them to be pair-wise (two-player game, Table \ref{2plpaytabex}) or in triplets (three-player game, \ref{3plpaytabex}).
We notice that the abundances of the three strains are qualitatively and quantitatively different in the two settings even though the underlying rules defining the interactions are the same.
Under neutrality the abundance of all the strains would be given by the dashed line.
The full lines are the analytical results obtained by solving Eq.\ \eqref{abundeq}, assuming $b=1.0$, $c_1 = 0.6$, $c_2 = 0.2$ and a population size of $N=30$ with selection intensity $\delta =0.003$.
}
\label{fig:3}
\end{figure*}

We calculate the abundances of the strains in these different settings, cf.\ Figure \ref{fig:3}. 
Even when there are almost no mutations there is a quantitative difference between the average frequencies of the strains.
For a higher mutation probability the difference is also qualitative.
While the freeloaders never pay a cost they have the highest abundance in the two-player setting for any mutation probability.
For the same reasoning but in three-player games we see that the abundance of the freeloaders falls below that of $T_2$ for a certain range of mutation probability.

\section{Discussion}

Public goods games are often used as examples of multi-player games.
In the beginning there were the cooperators and defectors.
Then came the punishers and then the loners \citep{hauert:2002te,szabo:2002te}.
Now we talk about second order punishers, pool and peer punishers \citep{sigmund:2010aa} and more.
Studying these systems for small mutation rates and arbitrary selection intensity is almost becoming standard \citep{fudenberg:2006ee,hauert:2007aa,hauert:2008bb,van-segbroeck:2009mi,sigmund:2010aa}.
In the limit of weak selection our method allows to find out which strategy is most abundant for arbitrary mutation rates.

Yet, another important aspect of most social dilemmas and many other biological examples is that they involve multiple players \citep{stander:1992aa,broom:2003aa,milinski:2008lr,levin:2009aa}.
\cite{antal:2009aa,antal:2009hc} have made use of the coalescence approach to characterize the mutation process under neutrality and then apply it under weak selection to two-player games with $n$ strategies ($n \times n$).
Here we extend the approach to $d$-player games with $n$ strategies.

We give an example for an $n \times n \times n$ game and derive the analogous expressions for abundances of the strategies for arbitrary mutation rates.
When we increase the number of players to $d$, the payoff matrix becomes a $d$ dimensional object.
We run into the problem of whether the order of players matters or not.
Either way this does not influence our results but notation-wise it is easier if the order of players does not matter.
Adding a new player adds a new index to the payoff values.
For calculating the abundance we need to assess Eq.\ \eqref{dplayers}.
For solving the two sums in Eq.\ \eqref{dplayers} we need to know the different combinations of choosing $d$ players and $d+1$ players from the neutral coalescent stationary state.

To illustrate the complexity of the situation take for example $s_4$.
This is the probability that four chosen individuals have the same strategy.
In \ref{coalescentapp} we have shown that deriving $s_4$ depends on $s_3$ which depends on $s_2$ in turn.
Hence in general to derive $s_{d+1}$, we need to know $s_{d}, s_{d-1}, s_{d-2} \ldots , s_2$.
In addition, for general $d$-player games we need quantities such as $\bar{s}_{d+1}$, probability that $d+1$ individuals chosen in the stationary state all have different strategies.
In the absence of mutations such probabilities as $\bar{s}_{d+1}$ vanish and we are left with only $s_{d+1}$, as mentioned in \cite{ohtsuki:2010bb}.
If $n<d$ then $\bar{s}_{d+1}$ is zero and hence the terms dependent on it need to be recalculated.
After recalculation the terms which are affected either vanish or are automatically adjusted such that the result can be written again in the form for $L_k$'s, $M_k$'s and $H_k$'s.
However, for a $d$ player game, $d-2$ intermediate terms such as $M_k$ appear.

For two strategies, $L_k$ reduces to the general condition derived in \citep{kurokawa:2009aa,gokhale:2010pn} and again holds for arbitrary $d$.
For $H_k$ for any number of strategies we conjecture that it will always be of the form $\frac{1}{n^{d}}\left[n \left(\sum_{r_2,\ldots r_d} a_{k,r_2,\ldots r_d } \right)-\left(\sum_{r_2,\ldots r_{d+1}} a_{r_2,\ldots r_{d+1} } \right)\right]$ from Eq.\ \eqref{dplayers}.
Addressing a generalization for $d$-player games is not a fundamental problem of the approach but requires a tedious 
recursive effort and poses a notational challenge.
At the coalescence level the problem rests on permutations and combinations.

Arbitrary mutation rates can be interpreted in different ways.
In the social learning context \citep{traulsen:2009aa} it can be thought of as the exploration rate where the players experiment with different strategies.
A recent behavioral experiment  suggests that humans change their strategies with a probability much higher than usually considered in theoretical models \citep{traulsen:2010pn}.
Analytical results for such high mutation rates can now be provided utilising the approach outlined herein.
Small mutation rates are most relevant in population genetic contexts where the strategies can be thought of as alleles.
While most people think of  evolutionary game theory as a phenotypic approach, 
one can as well consider evolutionary games on the level of genes \citep{hofbauer:1982mm,sigmund:1987aa,rowe:1987aa,rowe:1988aa,cressman:1992aa,hofbauer:1984mm,hofbauer:1998mm}.
A long standing problem in theoretical biology is to explain the maintenance of polymorphisms.
So far, the corresponding models typically assumed small mutation rates, but our approach provides a way to calculate the abundances for arbitrary mutation rates.
However, our aim is not to calculate abundances, but to depict a way to derive them.
A two-player interaction can capture the effect of one allele on another, but the epistatic interaction of many loci
calls for a theory of many-player interactions.

Making use of our approach we can precisely determine the composition of a population with any finite number of different types under weak selection for arbitrary mutation rates.
Another convenient way of finding the strategy which performs the best is the limit of small mutation rates.
For small mutation rates, the system spends most of its time in a monomorphic state.
We can approximate the system by just looking at fixation probabilities of the different types.
Our approach illustrates that the interaction of $d$ players is significantly more complex than the usual two-player games.
General multi-player games pose exciting challenges way beyond the usual intricacies of public goods games.\\

\textbf{Acknowledgements}. The authors would like to thank Bin Wu for helpful discussions and suggestions throughout the preparation of the manuscript.
The suggestions of two anonymous referees are highly appreciated.
C.S.G. and A.T. acknowledge support from the Emmy-Noether program of the Deutsche Forschungsgemeinschaft and from the Max-Planck Society.

\appendix

\section{Deriving the average change due to selection}
\label{eq1app}

We begin the Appendix by first deriving the average change in the frequency of a strategy under selection for an arbitrary number of players ($d$) and strategies ($n$).
In \ref{averagesapp}, we consider the special case $d=3$.
The more technical calculations based on coalescence theory can be found in Section \ref{coalescentapp}.

Let us begin with the simple case of a two-player game.
The payoff matrix for a two-player game, $\mathbf{A}$ with $n$ strategies is an $n \times n$ matrix,
\begin{eqnarray}
\label{2plnmat}
\mathbf{A}=
\bordermatrix{ 
   & 1 & 2 & \hdots & n \cr
1 & a_{1,1} & a_{1,2} & \hdots & a_{1,n} \cr
2 & a_{2,1} & a_{2,2} & \hdots & a_{2,n} \cr
\vdots & \vdots & \vdots & \ddots & \vdots \cr
n & a_{n,1} & a_{n,2} & \hdots & a_{n,n}
}
\end{eqnarray}
The average fitness of strategy $1$ can be written down as,
\begin{eqnarray}
\mathbf{f}_1 = 1 + \delta \left( \sum_{h=1}^{n} a_{1,h} x_h \right)
\end{eqnarray}
where the leading $1$ is the baseline fitness and $\delta > 0$ is the intensity of selection.
The variable $x_h$ is the frequency of players with strategy $h$, $\sum_{h=1}^{n} x_h = 1$.
We assume that the $\delta$ is so small that the fitness is always positive.
Similarly, for a three-player game the payoffs have an additional index.
Thus we can write the average fitness of strategy $1$ as,
\begin{eqnarray}
\mathbf{f}_1 = 1 + \delta \left( \sum_{h,i} a_{1,h,i} \left( x_h x_i \right) \right)
\end{eqnarray}
As usual, the sums run from $1$ to $n$, the number of strategies.
Continuing up to $d$ players we now consider $r_2 \ldots r_d$ the strategies of players $2 \ldots d$ as the strategy of one of the players is set to $k$, i.e. $r_1 = k$.
We see that the average payoff of strategy $k$ can be written as,
\begin{eqnarray}
\mathbf{f}_k = 1 + \delta \left( \sum_{r_2,\ldots, r_d} a_{k,r_2,\ldots, r_d } \left( x_{r_2} x_{r_3} \cdots x_{r_d} \right) \right)
\end{eqnarray}
The average payoff of the whole population is given by $\mathbf{F}$ as,
\begin{eqnarray}
\mathbf{F} = \sum_{k=1}^{n} x_{k}\mathbf{f}_{k}
\end{eqnarray}
Now we need to consider the dynamics of the process.
The Moran process is used, where in each time-step an individual is chosen proportional to its fitness to reproduce and a randomly chosen individual dies.
With probability $1-u$ the individual chosen for reproduction produces an exact copy as itself, but with probability $u$, a mutation occurs and the offspring can be of any of the $n$ strategies.

If the abundance of a strategy is greater than $1/n$, then it is favoured at the mutation-selection equilibrium.
To calculate the abundance of strategy $k$ we begin with the average number of offsprings of an individual of strategy $k$,
which is given by,
\begin{eqnarray}
\omega_k = 1 - \frac{1}{N} + \frac{1}{N} \frac{\mathbf{f}_k}{\mathbf{F}} .
\end{eqnarray}
The first term captures the survival of the parent.
The second and third terms refer to the random death and fitness proportional reproduction.
For $\delta \ll 1$, we have,
\begin{eqnarray}
\omega_k &\approx& 1+ \frac{\delta}{N} \Bigg[ \left( \sum_{r_2, \ldots, r_d} a_{k,r_2,\ldots, r_d } \left( x_{r_2} x_{r_3} \cdots x_{r_d} \right) \right) \nonumber \\
&& - \left ( \sum_{r_{d+1}} x_{r_{d+1}} \sum_{r_2,\ldots, r_d} a_{r_1,r_2,\ldots, r_d } \left( x_{r_2} x_{r_3} \cdots x_{r_d} \right) \right) \Bigg]. \nonumber \\
\end{eqnarray}
The change in the frequency of strategy $k$, $x_k$, due to selection is given by,
\begin{eqnarray}
\Delta x_k^{sel} = x_k \omega_k - x_k.
\end{eqnarray}
The vector $\mathbf{x} = (x_1, \ldots, x_n)$ contains all possible frequency compositions of the system.
The system will be in state $\mathbf{x}$ with probability $Q_\delta (\mathbf{x})$.
Hence by averaging $\Delta x_k^{sel}$ 	in the leading order of $\delta$ we obtain,
\begin{eqnarray}
\langle \Delta x_k^{sel} \rangle &\approx& \sum_{\mathbf{x}} \Delta x_k^{sel} Q_\delta (\mathbf{x}) \nonumber \\
&=& \delta \sum_{\mathbf{x}} \Bigg(\frac{1}{N} x_k \Bigg[ \left( \sum_{r_2,\ldots r_d} a_{k,r_2,\ldots r_d } \left( x_{r_2} x_{r_3} \ldots x_{r_d} \right) \right) \nonumber \\
&&- \left ( \sum_{r_2,\ldots r_{d+1}} a_{r_2,\ldots r_{d+1} } \left( x_{r_2} x_{r_3} \ldots x_{r_{d+1}} \right) \right) \Bigg] \Bigg) Q_\delta (\mathbf{x}) \nonumber \\
\end{eqnarray}
Thus we reach the expression for the average change in the frequency of strategy $k$ due to selection in the stationary state as,
\begin{eqnarray}
\langle \Delta x_k^{sel} \rangle_\delta &\!\!\!=& \frac{\delta}{N}  \langle x_k \Bigg[ \left( \sum_{r_2, \ldots r_d} a_{k,r_2,\ldots r_d } \left( x_{r_2} x_{r_3} \ldots x_{r_d} \right) \right) \nonumber \\
&& - \left ( \sum_{r_2,\ldots r_{d+1}} a_{r_2,\ldots r_{d+1} } \left( x_{r_2} x_{r_3} \ldots x_{r_{d+1}} \right) \right) \Bigg]  \rangle \nonumber \\
&\!\!\!=& \frac{\delta}{N}  \Bigg( \sum_{r_2,\ldots r_d} a_{k,r_2,\ldots r_d } \langle x_k x_{r_2} x_{r_3} \ldots x_{r_d} \rangle \nonumber \\
&&-  \sum_{r_2,\ldots r_{d+1}} a_{r_2,\ldots r_{d+1} } \langle x_k x_{r_2} x_{r_3} \ldots x_{r_{d+1}} \rangle \Bigg) \nonumber \\
\end{eqnarray}
Notice the form of a replicator like equation in the above terms.
We look for the difference between the average payoff of a strategy and the average payoff of the population.
The first sum consists of a product of $d$ frequencies while the second sum requires a product of $d+1$.
Particularly, we consider the case $d=3$.
For strategy $k$ the average change due to selection is given by
\begin{eqnarray}
\label{d3avg}
\langle \Delta x_k^{sel} \rangle_\delta = \frac{\delta}{N} \left(\sum_{h,i} a_{k,h,i}\langle x_k x_h x_i\rangle - \sum_{h,i,j} a_{h,i,j}\langle x_k x_h x_i x_j\rangle \right).\nonumber \\
\end{eqnarray}

\section{Three player games}
\label{averagesapp}

\subsection{Choosing a set of players}

To solve Eq. \eqref{d3avg}  we need to solve the two sums on the right hand side.
The first sum can be solved using the technique derived by \cite{antal:2009hc}.
For the second sum we need to know the different forms of the averages possible.
Using symmetry arguments such as $\langle x_1 x_2 x_2 x_3 \rangle = \langle x_1 x_2 x_3 x_3 \rangle$ (this is valid because we average under neutrality) only five different kinds of averages are required,
$\langle x_1 x_1 x_1 x_1 \rangle$, $\langle x_1 x_2 x_2 x_2 \rangle $, $\langle x_1 x_1 x_2 x_2 \rangle$, $\langle x_1 x_1 x_2 x_3 \rangle$ and $\langle x_1 x_2 x_3 x_4 \rangle$.
The quantities $\langle x_1 x_1 x_1 x_1 \rangle$ and $\langle x_1 x_2 x_3 x_4 \rangle$ are derived in section \ref{coalescentapp} based on coalescence theory.
The rest of the averages can be written down as,

\begin{enumerate}[(i)]
\item Three of a kind,  $\langle x_1 x_2 x_2 x_2\rangle$.
\begin{eqnarray}
\langle x_1 x_2 x_2 x_2\rangle &=& \left \langle \left( 1- \sum_{i=2}^n x_i\right) x_2 x_2 x_2 \right \rangle  \nonumber \\
&=& \langle x_1 x_1 x_1\rangle - \langle x_1 x_1 x_1 x_1\rangle - (n-2) \langle x_1 x_2 x_2 x_2 \rangle \nonumber \\
&=& \frac{ \langle x_1 x_1 x_1\rangle - \langle x_1 x_1 x_1 x_1\rangle}{n-1} 
\label{prob1}
\end{eqnarray}
\item Two pairs, $\langle x_1 x_1 x_2 x_2 \rangle$.
\begin{eqnarray}
\langle x_1 x_1 x_2 x_2 \rangle &=& \left \langle  (1-x_2- \sum_{i=3}^{n }x_i) x_1 x_2 x_2 \right \rangle \nonumber \\
&=& \langle x_1 x_2 x_2 \rangle - \langle x_1 x_2 x_2 x_2 \rangle - (n-2)\langle x_1 x_1 x_2 x_3 \rangle \nonumber \\
\label{prob2}
\end{eqnarray}
\item Single pair, $\langle x_1 x_1 x_2 x_3 \rangle$.
\begin{eqnarray}
\langle x_1 x_1 x_2 x_3 \rangle &=& \left \langle  \left(1- x_2 - x_3 - \sum_{i=4}^n x_i\right) x_1 x_2 x_3 \right \rangle \nonumber \\
&=&\langle x_1 x_2 x_3 \rangle - 2 \langle x_1 x_1 x_2 x_3 \rangle - (n-3) \langle x_1 x_2 x_3 x_4 \rangle \nonumber \\
&=& \frac{\langle x_1 x_2 x_3 \rangle - (n-3) \langle x_1 x_2 x_3 x_4 \rangle}{3}.
\label{prob3}
\end{eqnarray}
\end{enumerate}
Thus we can write all averages in terms of $\langle x_1 x_1 x_1 x_1 \rangle$, $\langle x_1 x_2 x_3 x_4 \rangle$ and the known quantities from \citep{antal:2009hc},
\begin{eqnarray}
\label{rawavgs}
\langle x_1 x_2 x_2 x_2\rangle &=& \frac{\langle x_1 x_1 x_1 \rangle - \langle x_1 x_1 x_1 x_1 \rangle }{n-1} \nonumber \\
\langle x_1 x_1 x_2 x_2 \rangle &=& \langle x_1 x_2 x_2 \rangle - \langle x_1 x_2 x_2 x_2 \rangle - (n-2) \langle x_1 x_1 x_2 x_3 \rangle \nonumber \\
\langle x_1 x_1 x_2 x_3 \rangle &=&  \frac{\langle x_1 x_2 x_3 \rangle - (n-3) \langle x_1 x_2 x_3 x_4 \rangle }{3} \nonumber \\
\end{eqnarray}
From \citep{antal:2009hc} we know the form of,
\begin{eqnarray}
\label{knownavgs}
\langle x_1 x_1 x_1\rangle &=& \frac{s_3}{n} \nonumber \\
\langle x_1 x_2 x_2\rangle &=& \frac{s_2 - s_3}{n (n-1)} \nonumber \\
\langle x_1 x_2 x_3\rangle &=& \frac{1- 3 s_2 + 2 s_3}{n (n-1) (n-2)},
\end{eqnarray}
where the probability that if we choose $i$ individuals from the stationary state of a neutral coalescent then all $i$ have the same strategy is $s_i$.
The quantities $s_2$ and $s_3$ have been previously derived in \citep{antal:2009hc}.
For completeness we repeat the derivation in Section \ref{coalescentapp}.
In Section \ref{coalescentapp} $s_4$ is calculated, which is the probability of choosing four individuals from the neutral stationary state and all have the same strategy.
If there are $n$ strategies then the probability that all four have strategy $1$ is $s_4/n$.
Thus $\langle x_1 x_1 x_1 x_1 \rangle = s_4/n$.
Similarly, the probability that all four have different strategies is $\bar{s}_4$.
The exact case when the first individual has strategy $1$ , second has $2$ , third has $3$ and the fourth has $4$ is just $\langle x_1 x_2 x_3 x_4 \rangle = \bar{s}_4/(n(n-1)(n-2)(n-3))$.
Using this information we can get the expression for all the five averages as,
\begin{eqnarray}
\label{savgs}
\langle x_1 x_1 x_1 x_1\rangle &=& \frac{s_4}{n} \nonumber \\
\langle x_1 x_2 x_2 x_2\rangle &=& \frac{s_3-s_4}{n (n-1)} \nonumber \\
\langle x_1 x_1 x_2 x_2 \rangle &=&  \frac{\bar{s}_4 + 3 s_4 - 8 s_3 + 6 s_2 -1}{3 n (n-1)} \nonumber \\
\langle x_1 x_1 x_2 x_3 \rangle &=& \frac{1 - 3 s_2 + 2 s_3 - \bar{s}_4}{3 n (n-1) (n-2)} \nonumber \\
\langle x_1 x_2 x_3 x_4 \rangle &=&  \frac{\bar{s}_4}{n (n-1) (n-2) (n-3)}.
\end{eqnarray}
The quantities $s_2$, $s_3$, $s_4$ and $\bar{s}_4$ are derived in Section \ref{coalescentapp}.
Substituting these values above yields,
\begin{eqnarray}
\label{newavgs}
\langle x_1 x_1 x_1 \rangle &=& n (n+\mu) (2 n+\mu) (3 + \mu) C \nonumber \\
\langle x_1 x_2 x_2 \rangle &=& n \mu (n+\mu) (3 + \mu) C \nonumber \\
\langle x_1 x_2 x_3 \rangle &=& n \mu^2 (3 + \mu) C \nonumber \\
\langle x_1 x_1 x_1 x_1\rangle &=& (n+\mu) (2 n+ \mu) (3 n+\mu) C \nonumber \\
\langle x_1 x_2 x_2 x_2\rangle &=& \mu (n+\mu) (2 n + \mu) C \nonumber \\
\langle x_1 x_1 x_2 x_2 \rangle &=&  \mu (n+\mu)^2 C \nonumber \\
\langle x_1 x_1 x_2 x_3 \rangle &=& \mu^2 (n+\mu) C \nonumber \\
\langle x_1 x_2 x_3 x_4 \rangle &=&  \mu^3 C.
\end{eqnarray}
where $C = \left[n^4 (1+\mu) (2+\mu) (3+\mu)\right]^{-1}$.

\subsection{Number of strategies with respect to the number of players}
\label{sums}

Now that we know the form of the averages, we can begin expanding the sums from Eq. \eqref{d3avg}, first for $d=3$ and for $n>3$.
Consider the first sum,
\begin{eqnarray}
\sum_{h,i} a_{k,h,i} \langle x_k x_h x_i\rangle &=&
\langle x_1 x_1 x_1 \rangle a_{k,k,k}  \nonumber \\
&&+ \langle x_1 x_2 x_2 \rangle \sum_{\substack{h,i \\ k \neq h = i \neq k \\ h = k, i \neq k \\ i = k, h \neq k }} a_{k,h,i}  \nonumber \\
&&+ \langle x_1 x_2 x_3 \rangle \sum_{\substack{h,i \\ k \neq h \neq i \neq k }}  a_{k,h,i}.
\end{eqnarray}
For the ease of notation we denote the co-efficients on the right hand side by $\alpha_1 = a_{k,k,k}$, $\alpha_2 = \sum_{\substack{h,i \\ k \neq h = i \neq k \\ h = k, i \neq k \\ i = k, h \neq k }} a_{k,h,i} $, $\alpha_3 = \sum_{\substack{h,i \\ k \neq h \neq i \neq k }}  a_{k,h,i}$.
Hence, we have,
\begin{eqnarray}
\label{newsumone}
\sum_{h,i} a_{k,h,i} \langle x_k x_h x_i\rangle =
\langle x_1 x_1 x_1 \rangle \alpha_1
+ \langle x_1 x_2 x_2 \rangle \alpha_2
+ \langle x_1 x_2 x_3 \rangle \alpha_3. \nonumber \\
\end{eqnarray}
Similarly, the second sum in Eq. \eqref{d3avg} becomes,
\begin{eqnarray}
\sum_{h,i,j} a_{h,i,j}  \langle x_k x_h x_i x_j\rangle &=&
\langle x_1 x_1 x_1 x_1 \rangle \beta_1
+   \langle x_1 x_2 x_2 x_2 \rangle \beta_2 \nonumber \\
&&+  \langle x_1 x_1 x_2 x_2 \rangle \beta_3
+ \langle x_1 x_1 x_2 x_3 \rangle \beta_4 \nonumber \\
&&+  \langle x_1 x_2 x_3 x_4 \rangle \beta_5
\end{eqnarray}
Note that $\beta_1 = \alpha_1$.
Substituting the expressions for the averages from Eqs. \eqref{newavgs},
\begin{eqnarray}
&&\frac{\sum_{h,i} a_{k,h,i} \langle x_k x_h x_i\rangle}{C} \nonumber \\
&&= n (n+\mu) (2 n+\mu) (3 + \mu) \alpha_1 
+ n \mu (n+\mu) (3 + \mu)  \alpha_2 \nonumber \\ 
&&\ \  + n \mu^2 (3 + \mu) \alpha_3 \nonumber \\
&&=
6 n^3 \alpha_1
+ n \left[2 n^{2} \alpha_1 + 3 n (3 \alpha_1+\alpha_2)\right] \mu  \nonumber \\
&&\ \ + n \left[ n (3 \alpha_1 + \alpha_2) + 3 (\alpha_1+\alpha_2+\alpha_3)\right] \mu^2 \nonumber \\
&&\ \  + n (\alpha_1+\alpha_2+\alpha_3) \mu^3
\end{eqnarray}
for the first sum.
For the second sum,
\begin{eqnarray}
&&\frac{\sum_{h,i,j} a_{h,i,j} \langle x_k x_h x_i x_j \rangle}{C} \nonumber \\
&=& (n+\mu) (2 n+ \mu) (3 n+\mu)  \alpha_1 +   \mu (n+\mu) (2 n + \mu) \beta_2 \nonumber \\
&&\ \ +  \mu (n+\mu)^2 \beta_3
 + \mu^2 (n+\mu)  \beta_4
+  \mu^3  \beta_5  \nonumber \\
&=& 6 n^3 \alpha_1
+ n^2 (11 \alpha_1 + 2 \beta_2 + \beta_3) \mu \nonumber \\
&&+ n (6 \alpha_1 + 3 \beta_2 + 2 \beta_3 + \beta_4) \mu^2 \nonumber \\
&&+ (\alpha_1+ \beta_2 + \beta_3 + \beta_4 + \beta_5) \mu^3.
\end{eqnarray}
Going back to our original Equation \eqref{d3avg} and organising it in powers of $\mu$, we obtain,
\begin{eqnarray}
&&\frac{N \langle \Delta x_k^{sel} \rangle_\delta}{\delta C} = \left(\sum_{h,i} a_{k,h,i}\langle x_k x_h x_i\rangle - \sum_{h,i,j} a_{h,i,j}\langle x_k x_h x_i x_j\rangle \right) \nonumber \\
&&= \underbrace{\frac{1}{n^3}[n^2 (2 \alpha_1 (n-1) + 3 \alpha_2 - 2 \beta_2 - \beta_3) ]}_{L_k} \mu n^3 + \nonumber \\
&& 
\underbrace{\frac{1}{n^3}[n \left(\left(3 n-3\right) \alpha_1 +\left(n+3\right) \alpha_2 +3 \alpha_3 - 3 \beta_2 - 2 \beta_3 - \beta_4\right)]}_{M_k} \mu^2 n^3 \nonumber \\
&&+ \underbrace{\frac{1}{n^3}[n \left(\alpha_1 + \alpha_2 + \alpha_3\right) - \left(\beta_1 + \beta_2 + \beta_3 + \beta_4 + \beta_5\right)]}_{H_k} \mu^3  n^3 \nonumber \\
\end{eqnarray}
Notice that the coefficients of the different orders of $\mu$ consist only of the number of strategies and the payoff values.
Denoting the coefficients of $\mu n^3$, $\mu^2 n^3$ and $\mu^3 n^3$ as $L_k$, $M_k$ and $H_k$ respectively, we get the following result
\begin{eqnarray}
\label{deltaresapp}
\langle \Delta x_k^{sel} \rangle_\delta
&=& \frac{\delta \mu (L_k + M_k \mu + H_k \mu^2)}{N n (1+\mu) (2+\mu) (3+\mu)}
\end{eqnarray}

Next we consider $d=3$ and $n=3$.
In this case the sums in Eq. \eqref{d3avg} are,
\begin{eqnarray}
\sum_{h,i} a_{k,h,i} \langle x_k x_h x_i\rangle &=& \langle x_1 x_1 x_1 \rangle \alpha_1 + \langle x_1 x_2 x_2 \rangle \alpha_2 + \langle x_1 x_2 x_3 \rangle \alpha_3 \nonumber \\
\end{eqnarray}
and
\begin{eqnarray}
\sum_{h,i,j} a_{h,i,j} \langle x_k x_h x_i x_j\rangle &=&
\langle x_1 x_1 x_1 x_1 \rangle \alpha_1
+   \langle x_1 x_2 x_2 x_2 \rangle \beta_2 \nonumber \\
&&+  \langle x_1 x_1 x_2 x_2 \rangle \beta_3 
+ \langle x_1 x_1 x_2 x_3 \rangle \beta_4. \nonumber \\
\end{eqnarray}
Thus $\bar{s}_4 =0$ and we do not have the term $\langle x_1 x_2 x_3 x_4 \rangle$.
This changes the averages, $\langle x_1 x_1 x_2 x_2 \rangle$ and $\langle x_1 x_1 x_2 x_3 \rangle$ as they were dependent on $\langle x_1 x_2 x_3 x_4 \rangle$ (see Eqs. \eqref{savgs}).
Eqs. \eqref{newavgs} do not change, but $\beta_5 = 0$.
Solving the two sums using these expressions and evaluating Eq. \eqref{d3avg}, specifically for $n=3$, 
\begin{eqnarray}
\frac{N \langle \Delta x_k^{sel} \rangle_\delta}{\delta C } &=& \left(\sum_{h,i}^3 a_{k,h,i}\langle x_k x_h x_i\rangle - \sum_{h,i,j}^3 a_{h,i,j}\langle x_k x_h x_i x_j\rangle \right) \nonumber \\
&=& 9 \left[4\alpha_1 + 3 \alpha_2- 2 \beta_2 - \beta_3 \right] \mu \nonumber \\
&&+ 3 \left[6 \alpha_1 +6 \alpha_2 + 3 \alpha_3 -3 \beta_2 - 2 \beta_3 - \beta_4 \right] \mu^2 \nonumber \\
 &&+ \left[3 \left(\alpha_1 + \alpha_2 + \alpha_3\right) - \left(\beta_1 + \beta_2 + \beta_3 + \beta_4 \right) \right] \mu^3 \nonumber \\
\end{eqnarray}
which can be written in the form of Eq. \eqref{deltaresapp}.

Finally for $d=3$ and $n=2$ the sums in Eq. \eqref{d3avg} consist only of the following terms,
\begin{eqnarray}
\sum_{h,i}^2 a_{k,h,i} \langle x_k x_h x_i\rangle &=& \langle x_1 x_1 x_1 \rangle \alpha_1 + \langle x_1 x_2 x_2 \rangle \alpha_2.\nonumber \\
\end{eqnarray}
and
\begin{eqnarray}
\sum_{h,i,j}^2 a_{h,i,j}  \langle x_k x_h x_i x_j\rangle &=& \langle x_1 x_1 x_1 x_1 \rangle \alpha_1+   \langle x_1 x_2 x_2 x_2 \rangle \beta_2 \nonumber \\
&&+  \langle x_1 x_1 x_2 x_2 \rangle \beta_3.
\end{eqnarray}
The form of the averages does not change from the general form given in Eqs. \eqref{savgs} except for $\langle x_1 x_1 x_2 x_2\rangle$ which depends on $\bar{s}_4$.
Due to $n=2$, $\bar{s}_4 = 0$ and also $\alpha_3 = \beta_5 = \beta_4 =0$.
For this special case thus, we have
\begin{eqnarray}
\frac{N \langle \Delta x_k^{sel} \rangle_\delta}{\delta C } 
&=& \left(\sum_{h,i}^2 a_{k,h,i}\langle x_k x_h x_i\rangle - \sum_{h,i,j}^2 a_{h,i,j}\langle x_k x_h x_i x_j\rangle \right) \nonumber \\
&=&4 \left[ 2\alpha_1 + 3 \alpha_2- 2 \beta_2 -  \beta_3\right] \mu \nonumber \\
&&+ 2 \left[ ( 3 \alpha_1 + 5 \alpha_2 - 3 \beta_2 - 2 \beta_3) \right] \mu^2 \nonumber \\
&&+ \left[2 \left(\alpha_1 + \alpha_2\right) - \left(\beta_1 + \beta_2 + \beta_3 \right) \right] \mu^3
\end{eqnarray}
which can be cast in the form of Eq. \eqref{deltaresapp}.

This case is actually very well studied.
For multiple players and two strategies it has been recently shown that the condition for
strategy $A$ replacing strategy $B$ with a higher probability simply depends on the sums of the payoff values of the two strategies \citep{kurokawa:2009aa,gokhale:2010pn}.
This result is valid for random matching of players and small mutation rates.
In our case, the condition for small mutation rates is obtained by checking the condition $L_k > 0$, i.e.,
\begin{eqnarray}
2\alpha_1 + 3 \alpha_2- 2 \beta_2 -  \beta_3 > 0. 
\end{eqnarray}
Inserting the definitions of $\alpha_1$, $\alpha_2$, $\beta_1$ and $\beta_2$ and rearranging leads to 
\begin{eqnarray}
2 a_{1,1,1}&&+a_{1,1,2}+a_{1,2,1} + 2 a_{1,2,2} \nonumber \\
&&>  2 a_{2,1,1} + a_{2,1,2} + a_{2,2,1} + 2 a_{2,2,2}
\end{eqnarray}
Under random matching we have $a_{1,1,2} = a_{1,2,1}$ and $a_{2,1,2}=a_{2,2,1}$.
Thus, this is equivalent to
\begin{eqnarray}
a_{1,1,1}+a_{1,1,2}+ a_{1,2,2} &>&  a_{2,1,1} + a_{2,1,2}+ a_{2,2,2}
\end{eqnarray}
which is the condition derived in \citep{kurokawa:2009aa,gokhale:2010pn}.
When we do the same analysis for $n=2$ and increasing $d$, and compare the $L_k$ for each $d$, we will find a general form of the condition already given in  \citep{kurokawa:2009aa,gokhale:2010pn} for small mutation rates.

\section{Calculating probabilities based on Coalescence Theory}
\label{coalescentapp}

In the coalescence approach, we take a sample from the present generation and look back in time with respect to the sample.
Consider two copies of a gene.
Sometime back in the past they come together to a common ancestor.
This means the lineages of the two copies ``coalesce" back in time.
In general if we have a sample of $d$ individuals from the present then sometime back the lineages of two of the individuals will coalesce and there will be $d-1$ individuals.
In all thus there will be $d-1$ coalescence events until we arrive at the most recent common ancestor, the root of the coalescent.
\cite{kingman:1982cc,kingman:1982aa,kingman:1982bb,kingman:2000fk}
showed that the mathematical process of joining lineages leading up to the common ancestor can be analytically understood.
He also showed that the coalescent encompassess a broad class of population dynamics models including Wright-Fisher and Moran processes.

There are three assumptions of the most basic coalescent theory \citep{wakeley:2008aa},
\begin{itemize}
\item The population is not subdivided or structured.
\item The population size remains constant over time.
\item Genetic differences have no effect on the fitness of an individual.
In our case this means that different strategies have the same fitness, the neutral case.
\end{itemize}

We follow the approach developed in \citep{antal:2009aa,antal:2009hc}.
In a neutral Moran process two individuals will have the same ancestor in one update step with probability $2/N^2$.
We use a continuous time limit by rescaling the time such that $\tau = t (2/N^2)$.
We determine the results for a large, but finite population size $N$.

The beauty of the coalescence process lies in the separation of the genealogical part and the mutation process.
This is due to the assumption of neutrality.
Mutations occur at the rate of $\mu/2$ where $\mu = N u$ and $u$ is the probability with which the offspring obtains any one of the $n$ strategies at random.
The mutation probability $u$ can range from $0$ to $1$, but when the mutation probability is $1$ then the strategies would oscillate.
Hence we rescale the mutation rate by $1/2$.
It has been shown by \citep{kingman:1982aa,kingman:1982bb} that when $N$ is large, the coalescent time is exponentially distributed as,
\begin{eqnarray}
\label{ctd}
f_{i}(\tau) = \binom{i}{2}e^{-\binom{i}{2} \tau}.
\end{eqnarray}
On each trajectory no mutation occurs in time $\tau$ with probability 
\begin{eqnarray}
\gamma = e^{-\frac{\mu}{2} \tau}.
\end{eqnarray}

\subsection{Calculation of ${s_2}$}

First  we repeat the derivations of \citep{antal:2009hc} for completeness of the process.
Also this will help simplify the terminologies used in the next subsection.
The quantity $s_2$ is the probability that two individuals chosen randomly in a neutral coalescent process have the same strategy.
According to the coalescent back in time there was a single common ancestor of the two chosen individuals.
Immediately after the ancestor split there were two individuals of the same type.
Thus the $s_2^{*}$ family consists of only one configuration, a pair of identical individuals.
From then onwards to $\tau_2$, mutations can play a role.
Hence, the probability that the two individuals drawn have identical strategies when at the $s_2^{*}$ family level they have identical strategies, is given by $s_2^{*[2]}$,
\begin{eqnarray}
s_2^{*[2]} (\tau) = \gamma^2 + \frac{2}{n} \gamma (1-\gamma) +  \frac{1}{n}(1-\gamma)^2.
\end{eqnarray}
The index $[2]$ in the superscript describes the composition of the configuration.
In this case denoting that both the individuals are of the same strategy.
The terms on the right hand side from first to last can be described as follows.
(i) None of the trajectories mutate and hence the individuals have identical strategies with probability $1$.
(ii) At least one mutation occurs on one of the trajectories and the chance that the new strategy is identical to the other is $1/n$. As there are two trajectories this can happen in two ways.
(iii) When both the trajectories mutate the first one gets some strategy with probability $1$ and the second also mutates to the same strategy with probability, $1/n$.

This has to be weighted by the probability that we begin with two identical individuals at the $s_2^{*}$ family level.
As this is the only possible configuration, the probability is $1$.
Further we also need to integrate with the coalescent time density (Eq. \eqref{ctd}) to finally get $s_2$ as,
\begin{eqnarray}
s_2 &=& 1 \int_{0}^{\infty} s_2^{*[2]}(\tau) f_2(\tau) d\tau \nonumber \\
&=& \frac{n+\mu}{n (1+\mu)}.
\end{eqnarray}
This is a the case of a $k$-allele Moran model with replacement \citep{ewens:2004qe}.
\subsection{Calculation of ${s_3}$}
Now we take a step further.
What is the probability that three randomly chosen individuals will have the same strategy ?
The distribution of coalescent times is given by the density function for the coalescent event for three individuals which is given by $f_3(\tau) = 3 e^{-3 \tau}$.

Similarly as above we first investigate the $s_3^*$ family.
At time $\tau_2$ in the coalescent tree, one of the two individuals splits.
Thus in the $s_3^*$ family, two individuals will always have identical strategies.
In all there can be only two configurations, all three are identical or two have the same strategy and the third differs.

We consider the two cases separately.
If all three individuals have the same strategy at the $s_3^*$ family level, then the probability that they have identical strategies after time $\tau$ is,
\begin{eqnarray}
s_3^{*[3]}(\tau) &=&  \gamma^3 + \frac{3}{n}\gamma^2 (1-\gamma)  + \frac{3}{n^2} \gamma (1-\gamma)^2  +  \frac{1}{n^2}(1-\gamma)^3.\nonumber \\
\end{eqnarray}
If two individuals have the same strategy and the third one is different at the $s_3^*$ family level, then the probability that they have identical strategies after time $\tau$ is given by,
\begin{eqnarray}
s_3^{*[2|1]} (\tau) &=& 0\ \gamma^3 + \frac{1}{n} \gamma^2 (1-\gamma) + \frac{3}{n^2} \gamma (1-\gamma)^2 + \frac{1}{n^2}(1-\gamma)^3 \nonumber \\
&=& \frac{1}{n} \gamma^2 (1-\gamma) + \frac{3}{n^2} \gamma (1-\gamma)^2 + \frac{1}{n^2}(1-\gamma)^3.
\end{eqnarray}
In the superscript the index $[2|1]$ denotes that two individuals are of the same strategy and one is of a different strategy.
In this case we see that the first term for all three trajectories not mutating vanishes.
This is because when we begin with the case when all the individuals do not have identical strategies, they cannot be identical later in time if no mutation occurs.

To get the full probability $s_3$ we need to weight the above two cases with the probabilities of their realizations.
Three individuals will be the same at the $s_3^*$ family level if the two individuals at $\tau_2$ are identical.
This happens with probability $s_2$.
The probability that they are not the same is thus $1-s_2$.
Putting in these weights and integrating over all possible times, we get $s_3$ as
\begin{eqnarray}
s_3 &=& s_2 \int_{0}^{\infty} s_3^{*[3]}(\tau) f_3(\tau) d\tau + (1- s_2) \int_{0}^{\infty} s_3^{*[2|1]}(\tau) f_3(\tau) d\tau \nonumber \\
&=& \frac{(n+\mu) (2 n+ \mu)}{n^2 (1+\mu) (2+\mu)}.
\end{eqnarray}

\subsection{Calculation of ${s_4}$}
\label{probs4}
Here we calculate $s_4$, i.e. the probability that four randomly chosen individuals have the same strategy out of a collection of $n$ strategies.

\begin{figure}
\includegraphics[width=\columnwidth]{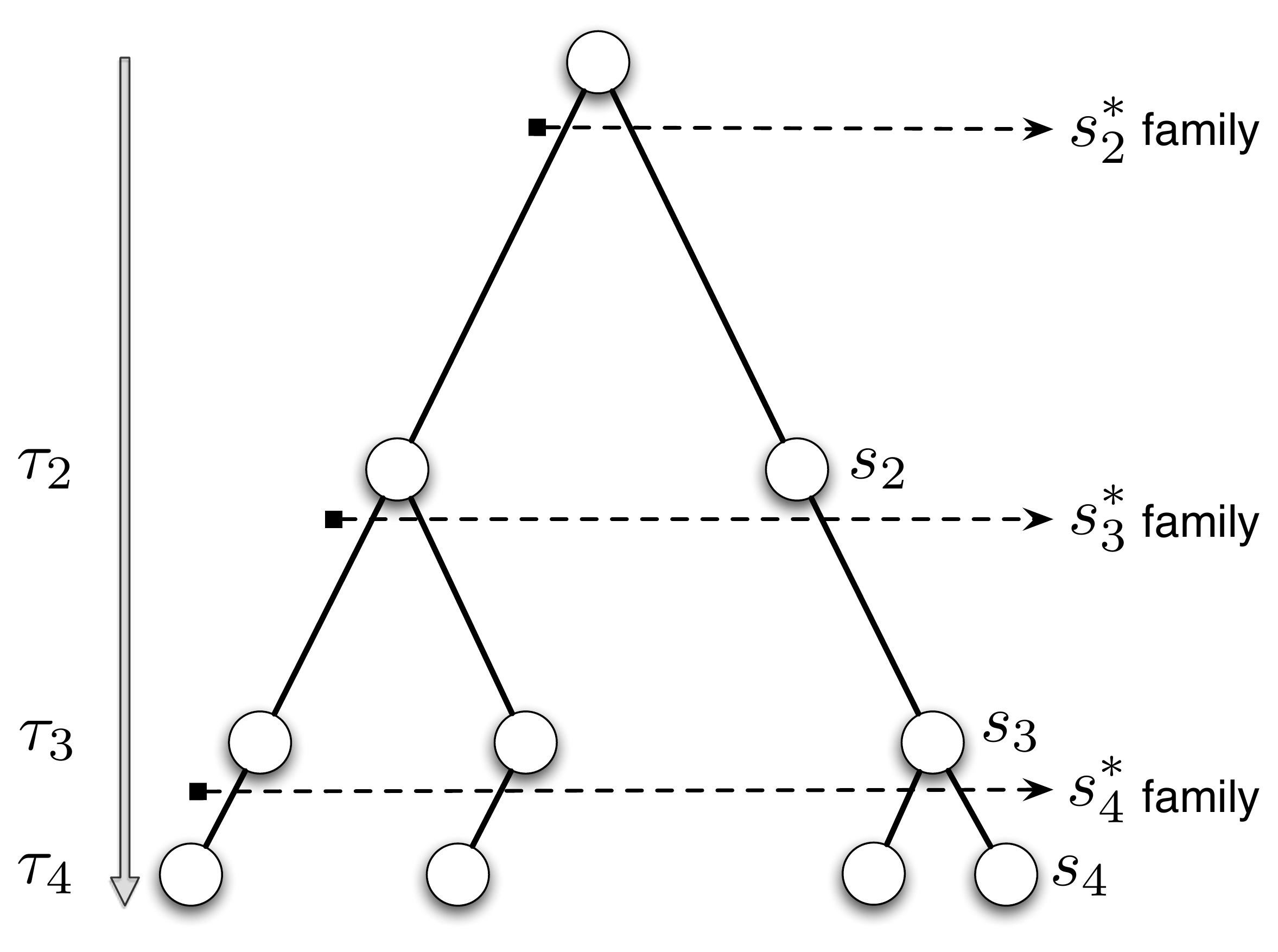}
\caption{
The coalescent as it evolves through time.
The probability that at time $\tau_i$ the $i$ individuals have the same strategy is given by $s_i$.
Immediately after $\tau_i$ there are $i+1$ individuals.
The strategy configuration at that time point depends if $s_i$ was $1$ or not.
If not then exactly what was the configuration?
All these factors determine the possible configuration of the immediate $i+1$ individuals and these different possibilities are grouped in the $s_{i+1}^*$ family.
}
\label{fig:c3}
\end{figure}

We are interested in the probability that the four leaves of the coalescent have the same strategy, cf. Fig. \ref{fig:c3}.
At time $\tau_3$, two of the four trajectories coalesce with rate 1.
Hence there is a coalescence at rate 6, and the density function is given as,
$f_{4}(\tau_4) = 6e^{-6 \tau_4}$.
Before the bifurcation occurs at $\tau_3$ the three players can have the same strategy with probability $s_3$ or at least one is different with probability $1-s_3$.

If the three players have the same strategy then immediately after the coalescence there will be four players with the same strategy.
If the three players do not have the same strategy then there are three different possible configurations.
This is the family of configurations we denote by $s_4^*$.
Thus beginning with four individuals of different configurations we are interested in the probability that after time $\tau$ all four of them will have the same strategy.

\begin{figure}
\includegraphics[width=0.5\columnwidth]{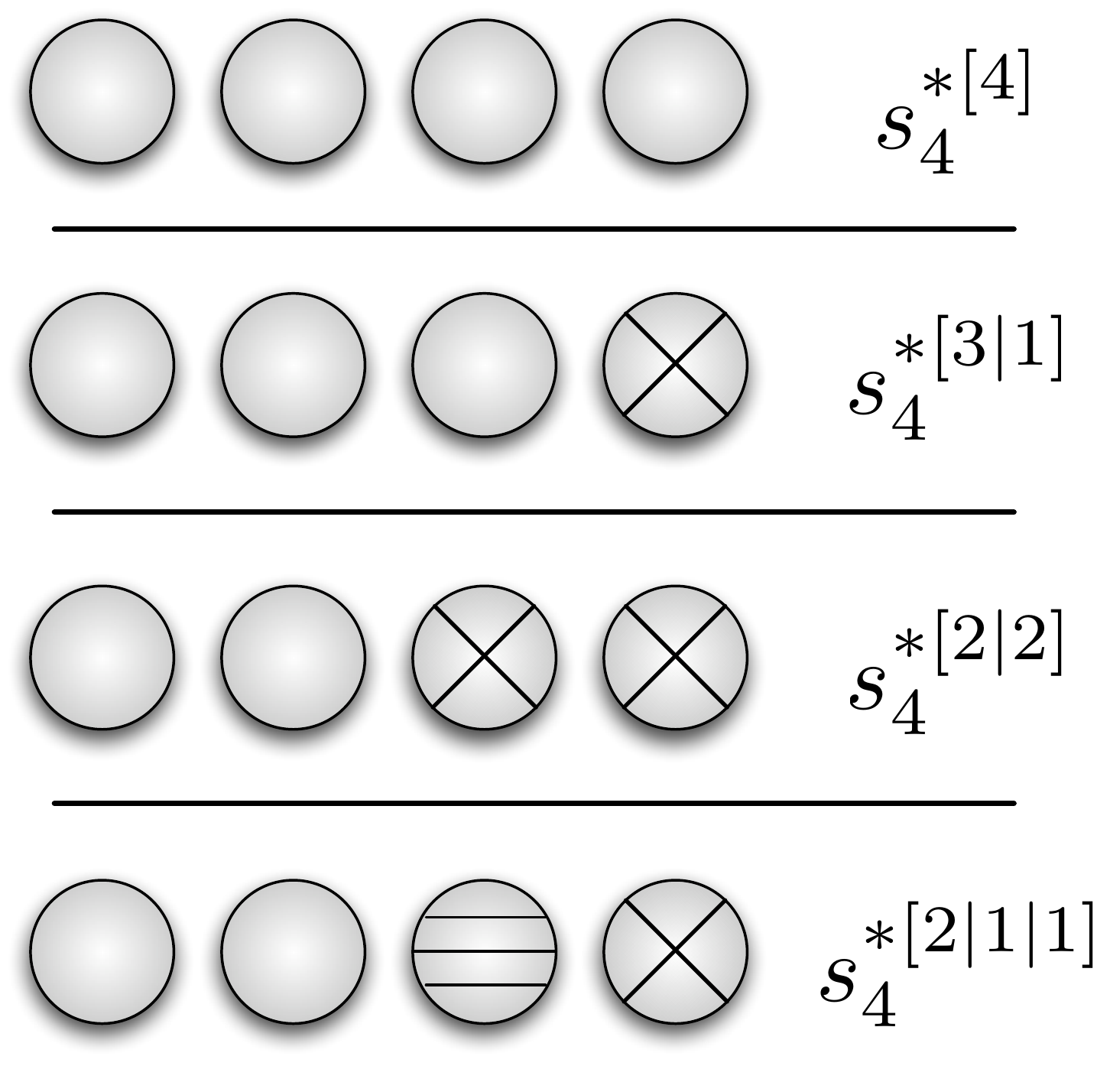}
\caption{
\textbf{The $s_4^*$ family.}
All possible starting configurations where there are $4$ individuals. Two of them have the same strategy.
The figure shows all the possible combinations for the remaining two individuals.
}
\label{fig:c4}
\end{figure}

The $s_4^*$ family consists of four cases:

\begin{itemize}
\item Four identical individuals (Fig. \ref{fig:c4}\ , $s_4^{*[4]}$). 
In this case they will be the same at time $\tau_4$ if none of them mutate.
If one of them mutates that can happen with probability $4 (\gamma)^3 (1-\gamma)$ and they are the same with probability $1/n$.
Similarly, we can write down when two or three or all can mutate and we get the expression,
\begin{eqnarray}
s^{*[4]}_4(\tau) &=& 
\gamma^4+\frac{4}{n}\gamma^3 (1-\gamma) +\frac{6}{n^2}\gamma^2 (1-\gamma)^2 \nonumber \\
&&+\frac{4}{n^3}\gamma (1-\gamma)^3+\frac{1}{n^3}(1-\gamma)^4.
\end{eqnarray}
\item Three of a kind (Fig. \ref{fig:c4}\ , $s_4^{*[3|1]}$).
If only three are the same then if no one mutates it is impossible for all four to be the same at time $\tau_4$.
Similarly we can argue what happens if one, two, three or all four mutate and we get the expression for $s_4^{*[3|1]}$:
\begin{eqnarray}
s^{*[3|1]}_4(\tau) &=& \frac{1}{n} \gamma^3 (1-\gamma) + \frac{3}{n^2} \gamma^2 (1-\gamma)^2 \nonumber \\
&&+ \frac{4}{n^3} \gamma (1-\gamma)^3 + \frac{1}{n^3}(1-\gamma)^4.
\end{eqnarray}
\item Two pairs (Fig. \ref{fig:c4}\ , $s_4^{*[2|2]}$).
At least two need to mutate such that we can end up with four identical individuals.
Additionally the two mutating must belong to the same pair.
The last two terms are the same as before:
\begin{eqnarray}
s_4^{*[2|2]} (\tau) = \frac{2}{n^2} \gamma^2 (1-\gamma)^2 + \frac{4}{n^3} \gamma (1-\gamma)^3 + \frac{1}{n^3}(1-\gamma)^4. \nonumber \\
\end{eqnarray}
\item Single pair (Fig. \ref{fig:c4}\ , $s_4^{*[2|1|1]}$).
At least two mutations are necessary for all four individuals to have the same strategy.
The two mutations have to be on the trajectory of the non-paired individuals.
Again the last two terms are the same as before:
\begin{eqnarray}
s_4^{*[2|1|1]} (\tau) = \frac{1}{n^2} \gamma^2 (1-\gamma)^2  + \frac{4}{n^3} \gamma (1-\gamma)^3 + \frac{1}{n^3}(1-\gamma)^4. \nonumber \\
\end{eqnarray}
\end{itemize}

To obtain the final probability $s_4$ (all four individuals have the same strategy), we combine all the above scenarios.
But we need to weight each of the scenarios with the probability of the realization of the starting configuration.
E.g. if the system reaches the state of all individuals having the same strategy from the second element of the $s_4^*$ family, i.e. $s_4^{*[3|1]}$, then we have to weight it by the probability of that configuration, three of the same type and one different, Fig \ref{fig:c4}.
This is possible if at $\tau_3$ we do not have all three of the same type, but they must be of one of the the type $\langle x_1x_2x_2 \rangle$ or $\langle x_2x_1x_2 \rangle$ or $\langle x_2x_2x_1 \rangle$.
Not only this, but the bifurcation should occur at one of the two identical types ($x_2$) and not the different type ($x_1$), the probability of which is $\frac{2}{3}$.
Thus we have to weight $s_4^{*[3|1]}$ by $\frac{2}{3} \times (s_2-s_3) \times 3$.
We calculate these weights for all the family members of $s_4^*$ and thus get an expression for $s_4$ as,
\begin{eqnarray}
\label{s4prob}
s_4 &=& s_3 \int_{0}^{\infty} s_4^*(\tau) f_4(\tau) d\tau + 2 (s_2-s_3) \int_{0}^{\infty} s_4^{*[3|1]}(\tau) f_4(\tau) d\tau \nonumber \\
&&+  (s_2 - s_3) \int_{0}^{\infty} s_4^{*[2|2]}(\tau) f_4(\tau) d\tau \nonumber \\
&&+ (1-3 s_2 + 2 s_3) \int_{0}^{\infty} s_4^{*[2|1|1]}(\tau) f_4(\tau) d\tau \nonumber \\
&=& \frac{(3 n + \mu) (2 n + \mu) (n+\mu)}{n^3 (1+\mu)(2+\mu)(3+\mu)}.
\end{eqnarray}
\subsection{Calculation of ${\bar{s}_4}$}
\label{probs4diff}

Here we calculate the probability $\bar{s}_4$ of picking four individuals in the stationary state all having different strategies.
As before we can have  four different starting configurations, the same as shown in Figure \ref{fig:c4}.

\begin{figure}[h]
\includegraphics[width=\columnwidth]{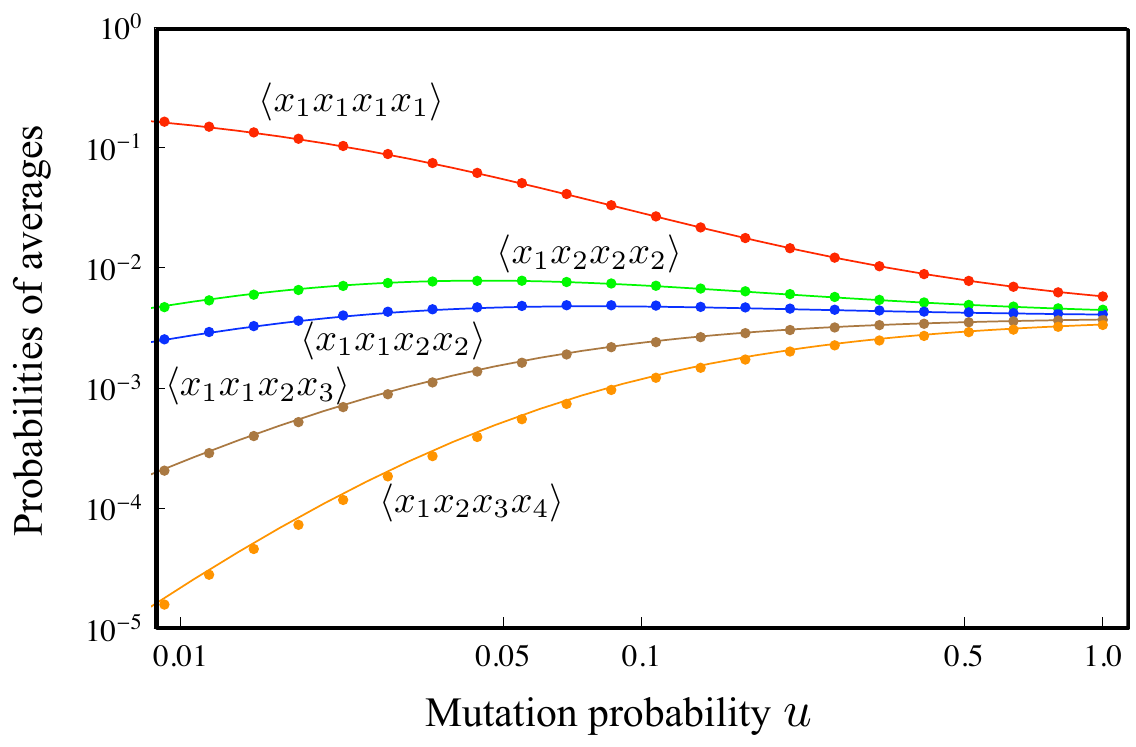}
\caption{
For a neutral Moran process with four strategies if we pick four individuals from the stationary state then the probability that all of them have the same strategy is given by, $s_4$, Eq. \eqref{s4prob}.
For four strategies ($n=4$), the probability that all four have strategy $1$ is $s_4/4$ given by $\langle x_1 x_1 x_1 x_1 \rangle$.
Similarly the probabilities for $\langle x_1 x_2 x_2 x_2 \rangle$, $\langle x_1 x_1 x_2 x_2 \rangle$, $\langle x_1 x_1 x_2 x_3 \rangle$ and $\langle x_1 x_2 x_3 x_4 \rangle$ are plotted as a function of the mutation probability for a population size of $N=40$.
The symbols are simulations while the lines are the analytical results.
}
\label{fig:c5}
\end{figure}

Hence basically now we want to calculate the probability that starting with each of the $s_4^*$ family members what is the probability of ending with all different individuals:

\begin{itemize}
\item Four identical individuals (Fig. \ref{fig:c4}\ , $\bar{s}_{4}^{*[4]}$).
We term the probability to start with four identical strategy individual to four different strategy individuals to be $\bar{s}_{4}^{*[4]}$.
For four to be different at least three have to mutate.
It can be calculated as follows,
\begin{eqnarray}
\bar{s}^{*[4]}_{4}(\tau) &=& 4 \gamma (1-\gamma)^3  \frac{(n-1) (n-2) (n-3)}{n^3} \nonumber \\
&&+ (1-\gamma)^4  \frac{(n-1) (n-2) (n-3)}{n^3}.
\end{eqnarray}
\item Three of a kind (Fig. \ref{fig:c4}\ , $\bar{s}_{4}^{*[3|1]}$).
For all four individuals to be different now we need at least two individuals to mutate as we already have one individuals of a different type.
Hence,
\begin{eqnarray}
\bar{s}^{*[3|1]}_4(\tau) &=&  3 \gamma^2 (1-\gamma)^2  \frac{(n-2) (n-3)}{n^2} \nonumber \\
&&+ 4 \gamma (1-\gamma)^3  \frac{(n-1) (n-2) (n-3)}{n^3} \nonumber \\ 
&& + (1-\gamma)^4  \frac{(n-1) (n-2) (n-3)}{n^3} . \nonumber \\
\end{eqnarray}
\item Two pairs (Fig. \ref{fig:c4}\ , $\bar{s}_4^{*[2|2]}$).
Here again we need at least two individuals to mutate for all the individuals to be different.
Of the two individuals mutating each should be of different types.
Hence, in all there are $4$ such combinations.
\begin{eqnarray}
\bar{s}_{4}^{*[2|2]} (\tau) &=& 4 \gamma^2 (1-\gamma)^2  \frac{(n-2) (n-3)}{n^2} \nonumber \\
&&+ 4 \gamma (1-\gamma)^3  \frac{(n-1) (n-2) (n-3)}{n^3} \nonumber \\
&&+ (1-\gamma)^4  \frac{(n-1) (n-2) (n-3)}{n^3}. \nonumber \\
\end{eqnarray}
\item Single pair (Fig. \ref{fig:c4}\ , $\bar{s}_4^{*[2|1|1]}$).
For this starting configuration a single mutation is enough to create all different individuals provided it happens in one of the paired individuals.
If two individuals are to mutate, then except for the two unpaired individuals together, all other groupings of two can give four different individuals, hence in $5$ different ways,
\begin{eqnarray}
\bar{s}_{4}^{*[2|1|1]} (\tau) &=& 2 \gamma^3 (1-\gamma) \frac{(n-3)}{n}+ 5 \gamma^2 (1-\gamma)^2 \frac{(n-2) (n-3)}{n^2} \nonumber \\
&&+ 4 \gamma (1-\gamma)^3  \frac{(n-1) (n-2) (n-3)}{n^3} \nonumber \\
&&+(1-\gamma)^4  \frac{(n-1) (n-2) (n-3)}{n^3}. 
\end{eqnarray}
\end{itemize}
To get the final probability $\bar{s}_4$ we need to integrate all the different starting configurations over the coalescent time density and add them all together.
Hence,
\begin{eqnarray}
\bar{s}_{4} &=& s_3 \int_{0}^{\infty} \bar{s}_{4}^{*}(\tau) f_4(\tau) d\tau
+ 2 (s_2-s_3) \int_{0}^{\infty} \bar{s}_{4}^{*[3|1]}(\tau) f_4(\tau) d\tau  \nonumber \\
&& + (s_2 - s_3) \int_{0}^{\infty} \bar{s}_{4}^{*[2|2]}(\tau) f_4(\tau) d\tau \nonumber \\
&&+ (1-3 s_2 + 2 s_3) \int_{0}^{\infty} \bar{s}_{4}^{*[2|1|1]}(\tau) f_4(\tau) d\tau \nonumber \\
&=& \frac{ \mu^3 (n - 1) (n - 2) (n - 3) }{n^3 (1+\mu)(2+\mu)(3+\mu)}.
\end{eqnarray}
Due to the notational challenge, possible errors can arise hence to check our results, we simulated a neutral Moran process and computed the different averages, $\langle x_1 x_1 x_1 x_1 \rangle$, $\langle x_1 x_2 x_2 x_2 \rangle$, $\langle x_1 x_1 x_2 x_2 \rangle$, $\langle x_1 x_1 x_2 x_3 \rangle$ and $\langle x_1 x_2 x_3 x_4 \rangle$.
These quantities depend on all the probabilities calculated in the Appendix namely $s_2$, $s_3$, $s_4$ and $\bar{s}_4$.
The results of the simulation and analytical method are shown in Figure \ref{fig:c5}.

\bibliographystyle{elsarticle-harv}

\end{document}